\documentclass[preprint,proceedings]{rmaa}
\usepackage{epsfig}
\usepackage{paralist}
\usepackage{psfrag,color}

\SetYear{2005}
\SetConfTitle{The Ninth Texas-Mexico Conference on Astrophysics}

\title{Helium Recombination Lines
as a Probe of Abundance and Temperature Problems}

\author{
 Y. Zhang,\altaffilmark{1,2}
 R. H. Rubin,\altaffilmark{3,4}
 and X.-W. Liu\altaffilmark{1}}

\altaffiltext{1}{Peking University, Beijing, P. R. China.}
\altaffiltext{2}{Space Telescope Science Institute, Baltimore, USA.}
\altaffiltext{3}{NASA Ames Research Center, Moffett Field, USA.}
\altaffiltext{4}{Orion Enterprises, M.S. 245-6, Moffett Field, USA.}

\shortauthor{Zhang, Rubin, \& Liu}
\shorttitle{He~{\sc i} lines}

\fulladdresses{
\item Y. Zhang: Department of Astronomy, Peking University, Beijing 100871,
       P. R. China; Space Telescope Science Institute,
3700 San Martin Drive, Baltimore, MD 21218, USA
 (\email{zhangy@stsci.edu}).
\item R. H. Rubin: NASA Ames Research Center, Moffett Field, CA 94035-1000, USA.
(\email{rubin@cygnus.arc.nasa.gov})
\item X.-W. Liu: Department of Astronomy, Peking University, Beijing 100871,
       P. R. China.(\email{liuxw@bac.pku.edu.cn})
}

\listofauthors{Y. Zhang, R. H. Rubin, \& X.-W. Liu}
\indexauthor{Zhang, Y}
\indexauthor{Rubin, R. H.}
\indexauthor{Liu. X.-W.}

\abstract{ The paper presents a simplified formula to
determine an electron temperature,
$T_{\rm e}$(\ion{He}{1}),
for planetary nebulae (PNe) using
the He~{\sc i} $\lambda7281/\lambda6678$ line
flux ratio.
In our previous studies
of $T_{\rm e}$(\ion{He}{1}) (Zhang et al. 2005), we used the
\ion{He}{1} line emission coefficients given by Benjamin
et al.
(1999). Here we examine the results of using
more recent atomic data presented by Porter et al. (2005).
A good agreement is shown, suggesting that the effect
of uncertainties of atomic data on the resultant $T_{\rm e}$(\ion{He}{1})
is negligible. We also present an
analytical formula
to derive electron temperature using the \ion{He}{1} discontinuity at
3421~{\AA}. Our analysis shows that $T_{\rm e}$(\ion{He}{1})
values are significantly
lower than electron temperatures
deduced from the Balmer jump
of \ion{H}{1} recombination
spectra, $T_{\rm e}$(\ion{H}{1}),
and that inferred from the collisionally excited [\ion{O}{3}]
nebular-to-auroral forbidden line
flux ratio,
$T_{\rm e}$([\ion{O}{3}]).
In addition, $T_{\rm e}$(\ion{H}{1})
covers a wider range of values than
either $T_{\rm e}$(\ion{He}{1})
or
$T_{\rm e}$([\ion{O}{3}]). This supports the two-abundance nebular model
with hydrogen-deficient material embedded in diffuse gas of a ``normal''
chemical composition (i.e.  $\sim$ solar).
}

\addkeyword{ISM: General (planetary nebula)}

\begin{document}
\maketitle

\section{Introduction}
\label{sec:intro}

% First sentence is way too long.
There are two long-standing problems in nebular
astrophysics that are termed the
``abundance problem" and the ``temperature problem".
The abundance problem refers
to the findings that when the heavy element abundances
are measured, particularly in planetary nebulae (PNe),
the results derived from collisionally excited lines (CEL) are often
lower than those derived from optical recombination  lines (ORLs).
One manifestation of the temperature problem is that
electron temperatures deduced from the collisionally
excited [\ion{O}{3}] nebular-to-auroral forbidden line ratio -- hereafter
$T_{\rm e}$([\ion{O}{3}]) -- are systematically higher than those determined
from the Balmer jump (BJ) of \ion{H}{1} recombination spectrum -- hereafter
$T_{\rm e}$(\ion{H}{1}) (see Liu 2004 for a recent review).
Some recent attempts to solve these problems have
implications that PNe may have complex physical conditions which are far
from being understood.  
A correlation between the  abundance discrepancies
and the temperature discrepancies is found by Liu et al.
(2001), suggesting that the two problems may have a common
origin.  Two possible solutions are
{\bf a)} the presence of temperature and density variations within chemically homogeneous nebulae
(Peimbert 1967; Viegas \& Clegg 1994)
and
{\bf b)} a two-abundance nebular model with hydrogen-deficient
material embedded in diffuse gas of a ``normal'' chemical composition
(i.e.  $\sim$ solar) (Liu et al. 2000). To further investigate the problems,
more methods to probe nebular physical conditions are obviously
valuable.

In an earlier paper (Zhang et al. 2005, hereafter Z05), we presented a
method to use He~{\sc i}
recombination lines to measure electron
temperatures of PNe -- hereafter
$T_{\rm e}$(\ion{He}{1}).
% removed square brackets
We studied the effect that temperature
and density variations inside nebulae have on the $T_{\rm e}$(\ion{He}{1}) value
and found that for a chemically homogeneous nebula, the possible presence of
temperature and density variations causes
$T_{\rm e}$(\ion{He}{1})$\ga$$T_{\rm e}$(\ion{H}{1}).
In contrast, the two-component nebular model predicts
$T_{\rm e}$(\ion{He}{1})$<$$T_{\rm e}$(\ion{H}{1}). Therefore,
a comparison between $T_{\rm e}$(\ion{He}{1}) and $T_{\rm e}$(\ion{H}{1})
provides an opportunity to discriminate between the two paradigms.
Applying the method to a sample of PNe,
we found that $T_{\rm e}$(\ion{He}{1}) values are
significantly lower than $T_{\rm e}$(\ion{H}{1}) values,
in agreement with the prediction of
the two-component (also called two-abundance) nebular model.

This paper is an extension of Z05. In Section~2, we present
an analysis formula to derive $T_{\rm e}$(\ion{He}{1}).
In Section~3, we give a statistical discussion of
$T_{\rm e}$(\ion{He}{1}), $T_{\rm e}$(\ion{H}{1})
and $T_{\rm e}$(\ion{O}{3}).

\section{Method}
%\label{sec:errors}

In order to derive $T_{\rm e}$(\ion{He}{1}),
we use analytic formulae for the emissivities of \ion{He}{1}
lines
as a function of electron temperature given by Benjamin et al. (1999).
We suggested
that the \ion{He}{1} $\lambda7281/\lambda6678$ intensity
ratio serves as the best line ratio suitable for temperature
determinations (Z05). A non-linear equation,
$I(7281)/I(6678)=f(T_{\rm e})$,
was used in Z05 to determine
$T_{\rm e}$(\ion{He}{1}).

However, we find that when $T_{\rm e}<15000$\,K,
there is a simple linear relationship
between $T_{\rm e}$(\ion{He}{1})
and the \ion{He}{1} $\lambda7281/\lambda6678$
intensity ratio, as shown in Fig.~1.
With an electron density $N_{\rm e}=10^4$\,cm$^{-3}$,
a least-squares fit yields the following relation
\begin{equation}
T_{\rm e}({\rm He~I })=49300\times \frac{I(7281)}{I(6678)}\ -2150~ {\rm K}.
\end{equation}
The relation is quite insensitive to electron density.
We thus recommend
equation~1 as a general formula to determine
$T_{\rm e}$(\ion{He}{1}) for $T_{\rm e}<15000$\,K.
Considering the uncertainties of electron density, the maximum
systematical errors
of $T_{\rm e}$(\ion{He}{1}) derived by
equation~1  is 7$\%$.
In Fig.~2 we compare the $T_{\rm e}$(\ion{He}{1}) derived from
equation~1  and those derived by Z05
using the non-linear equation mentioned above.
An excellent agreement is indicated.

Recently, based on improved atomic data, Porter et al. (2005) presented
new calculations
of \ion{He}{1}
line emissivities. They claimed that
for
the 32
\ion{He}{1} emission lines they considered,
the average difference
between the new \ion{He}{1} emissivities and those of Benjamin et al. (1999)
is 4.6$\%$. In Fig.~1 we overplot the values obtained according
to Porter et al. (2005) (cf. their Table~1).
A good
agreement between
the two sets of
data is shown, implying that there is hardly
any effect of the
uncertainties of \ion{He}{1} emissivities on
the temperature determination from the \ion{He}{1} $\lambda7281/\lambda6678$
intensity ratio.

The determination of $T_{\rm e}$(\ion{He}{1}) is based on the assumption of Case B
for the \ion{He}{1} lines. An important consideration, therefore, is whether
the possible departure from pure
Case B to Case A recombination for the \ion{He}{1} singlet lines
may affect temperature determination. In Z05, we also
used the
He~{\sc i} $\lambda7281/\lambda5876$ intensity ratios to determine
electron temperatures, which are found to be in good agreement with those
derived from the He~{\sc i} $\lambda7281/\lambda6678$ ratios.
Given the argument that the deviation from Case B has
a different
effect on
the He~{\sc i} $\lambda7281$ and the He~{\sc i} $\lambda5876$ lines,
we suggest that the
effect is negligible.

\begin{figure}
\epsfig{file=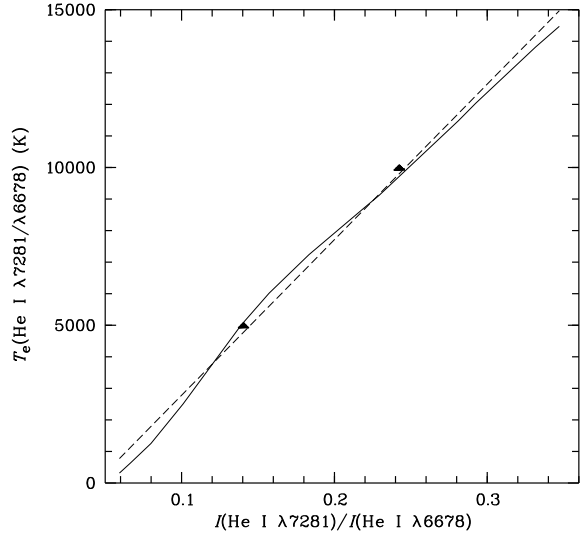,
height=7.2cm, bbllx=77, bblly=271, bburx=530, bbury=692, clip=, angle=0}
\caption{Electron temperature is plotted against the \ion{He}{1}
$\lambda7281$/$\lambda6678$ intensity ratio for
an electron
density of 10$^4$\,cm$^{-3}$.
The solid line shows the prediction based
on the atomic data
given by  Benjamin et al. (1999).
The dashed line is a linear fit (see equation~1).
The triangles represent the
prediction based on the more recent calculation of Porter et al. (2005).}
\label{mex1}
\end{figure}

\begin{figure}
\epsfig{file=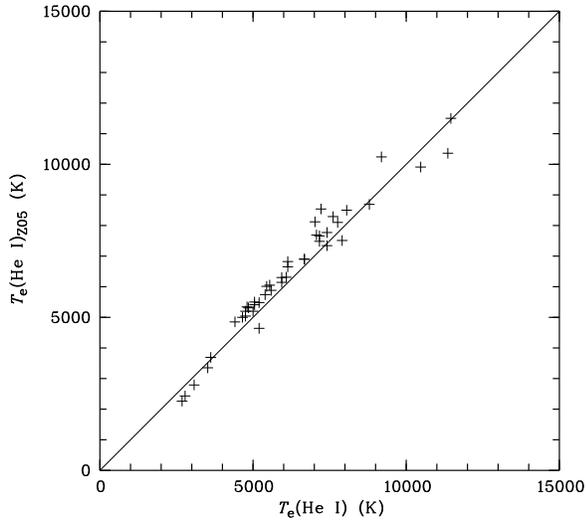,
height=7.2cm, bbllx=77, bblly=258, bburx=546, bbury=692, clip=, angle=0}
\caption{Comparison of the $T_{\rm e}$(\ion{He}{1}) derived from
equation~1 and those derived by Z05. The solid line is a $y=x$ plot.}
\label{mex1}
\end{figure}

Another method to measure $T_{\rm e}$(\ion{He}{1}) is  to use
the \ion{He}{1}
discontinuity at 3421\,{\AA}, produced by He$^{+}$ recombinations to the
\ion{He}{1} 2p\,$^3$P$^{\rm o}$ level. A relation between the
\ion{He}{1} discontinuity at 3421\,{\AA} and
electron temperature is derived as,
\begin{equation}
% T_{\rm e}({\rm He~I})=(2267\pm46)\times\left[\frac{J(3421)}{I(3613)}\right]^{-3/2}~K,
%  J usually means something else (average intensity).
T_{\rm e}({\rm He~I})=(2267\pm46)\times\left[\frac{\Delta I(3421)}{I(3613)}\right]^{-3/2}~K,
\end{equation}
% where $J(3421)/I(3613)=[I(\lambda3421^-)-I(\lambda3421^+)]/I({\rm He~I}~\lambda3613)$.
where
$\Delta I(3421)/I(3613)=[I(\lambda3421^-)-I(\lambda3421^+)]/I({\rm He~I}~\lambda3613)$.
The observation of the \ion{He}{1} $\lambda$3421 discontinuity,
however, is very difficult due to its weakness.
In this paper, we
present only a fit relation between this discontinuity
and electron temperature. The application of this method to PNe is beyond the
scope of this paper.
We note that to apply this method,
high quality spectroscopic observations of this discontinuity
are required.

\section{Results and discussion}

We calculate $T_{\rm e}$(\ion{He}{1}) for 48 PNe.
Our results
show that
$T_{\rm e}$(\ion{He}{1})$\la$$T_{\rm e}$(\ion{H}{1})$\la$$T_{\rm e}$([\ion{O}{3}]).
The average values for the sample PNe are
$T_{\rm e}$(\ion{He}{1})$=6300\pm2100$~K,
$T_{\rm e}$(\ion{H}{1})$ =10300\pm3100$~K, and
$T_{\rm e}$([\ion{O}{3}])$ =11900\pm2600$~K.
The result is exactly opposite to the predictions of the scenarios of
temperature fluctuation and density inhomogeneities but in good agreement
with the expectations of the two-abundance nebular model
proposed by Liu et al. (2000).

Fig.~3 shows the distribution of $T_{\rm e}$(\ion{He}{1}),
$T_{\rm e}$(\ion{H}{1}) and $T_{\rm e}$([\ion{O}{3}]).
Inspection of Fig.~3 shows that except for a few extreme cases,
$T_{\rm e}$([\ion{O}{3}]) falls between 9000 and 15\,000~K, and
$T_{\rm e}$(\ion{He}{1}) falls between 3000 and 9000~K,
whereas
$T_{\rm e}$(\ion{H}{1})
covers a wider range, varying from 5000 to 15000~K.
In the scenario of
the two-component nebular model, the
H-deficient component is extremely cold so that no CELs arise but
ORLs are heavily enhanced. Therefore,
$T_{\rm e}$(\ion{He}{1}) and
$T_{\rm e}$([\ion{O}{3}]) characterize the H-deficient cold gas and
the diffuse hot gas, respectively.  Due to its hydrogen deficiency,
the ultra-cold ionized gas has
much less of a
contribution to the \ion{H}{1}
recombination spectrum than the \ion{He}{1} recombination spectrum.
Accordingly, $T_{\rm e}$(\ion{H}{1})
is a weighted average over both regions.
As a result, $T_{\rm e}$(\ion{H}{1}) covers
a wider range of values than
$T_{\rm e}$(\ion{He}{1})
or $T_{\rm e}$([\ion{O}{3}]).

\begin{figure}
\epsfig{file=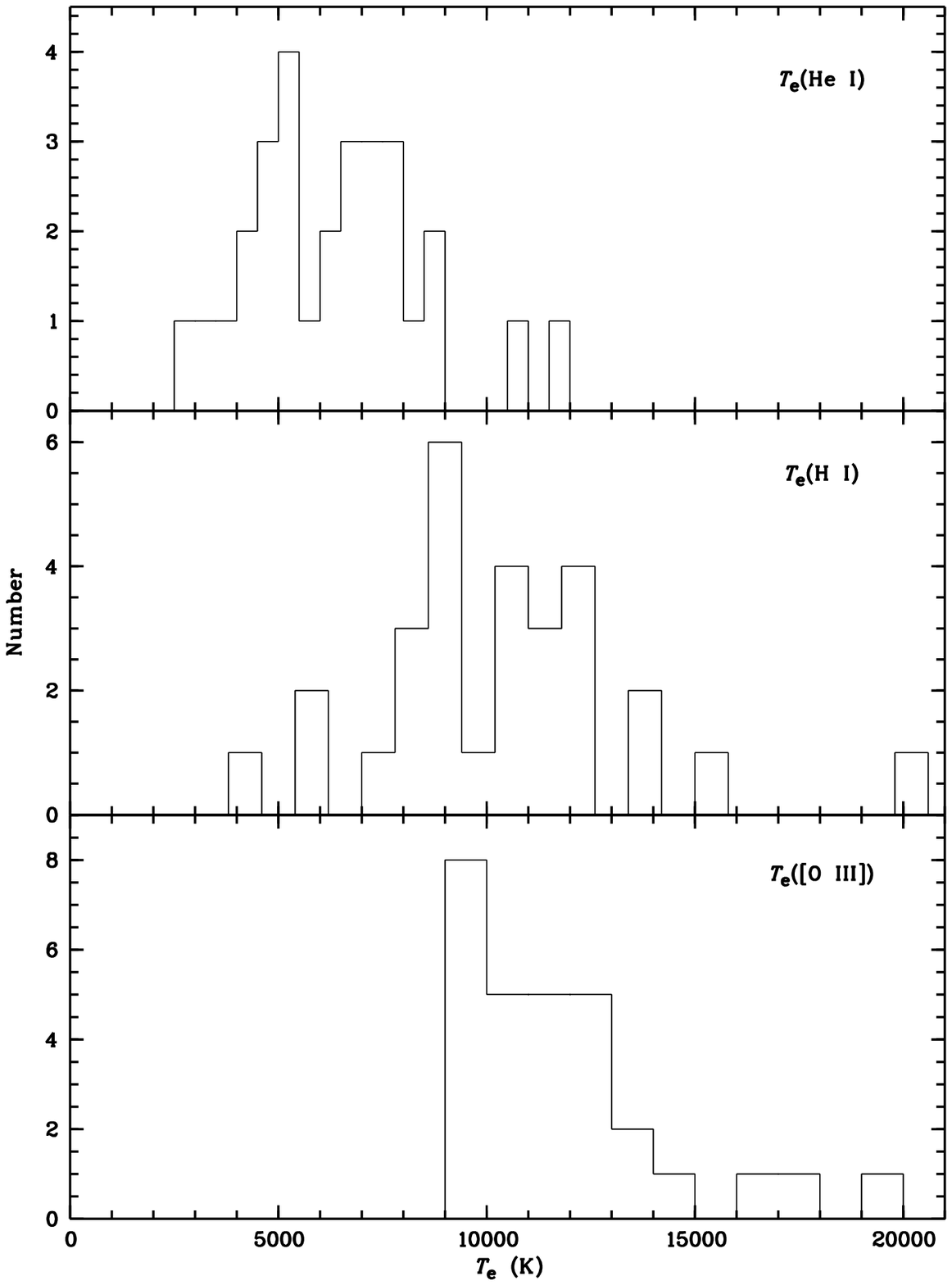,
height=9.6cm, bbllx=113, bblly=56, bburx=515, bbury=605, clip=, angle=0}
\caption{Histograms showing the numbers of PNe with a given
$T_{\rm e}$(\ion{He}{1}) (upper panel), $T_{\rm e}$(\ion{H}{1})
(middle panel) and
$T_{\rm e}$([\ion{O}{3}]) (lower panel).}
\label{mex3}
\end{figure}

In previous studies of a large sample
of PNe (Zhang et al. 2004),
we have found that the discrepancies between $T_{\rm e}$(\ion{H}{1})
and $T_{\rm e}$([\ion{O}{3}]) are anticorrelated with electron
densities, i.e. high-density nebulae have the smallest temperature
discrepancies. Robertson-Tessi
\& Garnett (2005) found a negative
correlation between the CEL/ORL abundance discrepancies and
the nebula diameter and the Balmer surface brightness, suggesting
that bright and compact PNe have small CEL/ORL abundance discrepancies.
Given that the abundance discrepancies are positively correlated with
the temperature discrepancies (Liu et al. 2001), the two findings are
mutually
consistent. We thus infer that the abundance and
temperature discrepancies are related to nebular evolution;
with expansion of nebulae, the contribution from the H-deficient
material to recombination
spectra
becomes increasingly important. Here we compare
the discrepancies between $T_{\rm e}$(\ion{He}{1})
and $T_{\rm e}$(\ion{H}{1}) against nebular densities and
CEL/ORL abundance discrepancies. However, no prominent correlation is seen,
implying that some properties of the postulated H-deficient
inclusions which dominate $T_{\rm e}$(\ion{He}{1}), such as its
helium composition, are very different for
different PNe. To further understand  the H-deficient
inclusions,  high S/N ratio, high spectral resolution
spectroscopy
of ORLs from heavy elements is required.

The presence of H-deficient inclusions may  have
an important
influence on the determination
of the He/H abundance.
Based on a large sample of PNe, Z05 estimated that the filling factor
of H-deficient components has a typical value of $10^{-4}$, which
may cause
the He$^+$/H$^+$ ionic abundance ratio
to be
overestimated by a factor of
$\sim$1.25.
The estimation, however, is
based on a very simple assumption,
that the H-deficient inclusions
in all PNe have identical properties
to those given by
P\'{e}quignot et al. (2003) in constructing
two-abundance photoionization models of NGC~6153.
Further spectroscopy of PNe should provide more
constraints on
the physical conditions of the H-deficient inclusions.

\acknowledgements

YZ acknowledges the award of an Institute Fellowship from STScI.
The work of YZ and XWL has been supported partially by Chinese NSFC Grant
NO.10325312.  Support for RHR
comes from the NASA Long-Term Space
Astrophysics (LTSA) program.

\end{document}